\documentclass[aps,draft,superscriptaddress,preprint,endfloats]{revtex4}

\usepackage{bm,graphicx,amsmath,amssymb,graphics}
\usepackage{textcomp}

\begin{document}

\preprint{}

\title{Electric-field-induced spin-flop in $BiFeO_3$ single crystals at room-temperature}
\author{D. Lebeugle}
\affiliation{Service de Physique de l'Etat Condens{\'e}, CEA
Saclay, F-91191 Gif-Sur-Yvette}
\author{D. Colson}
\affiliation{Service de Physique de l'Etat Condens{\'e}, CEA
Saclay, F-91191 Gif-Sur-Yvette}
\author{A. Forget}
\affiliation{Service de Physique de l'Etat Condens{\'e}, CEA
Saclay, F-91191 Gif-Sur-Yvette}
\author{M. Viret}
\affiliation{Service de Physique de l'Etat Condens{\'e}, CEA
Saclay, F-91191 Gif-Sur-Yvette}
\author{A. M. Bataille}
\affiliation{Laboratoire Leon Brillouin, CEA Saclay, F-91191
Gif-Sur-Yvette}
\author{A. Gukasov}
\affiliation{Laboratoire Leon Brillouin, CEA Saclay, F-91191
Gif-Sur-Yvette}

\begin{abstract}

Bismuth ferrite, $BiFeO_3$, is the only known room-temperature 'multiferroic' material. We demonstrate here, using neutron scattering measurements in high quality single crystals, that the antiferromagnetic and ferroelectric orders are intimately coupled. Initially in a single ferroelectric state, our crystals have a canted antiferromagnetic structure describing a unique cycloid. Under electrical poling, polarisation re-orientation induces a spin flop. We argue here that the coupling between the two orders may be stronger in the bulk than that observed in thin films where the cycloid is absent.

\end{abstract}


\maketitle

Electricity and magnetism are properties which are closely linked to each other. This link is dynamic in essence, as moving charges generate a magnetic field and a changing magnetic field produces an electric field. This forms the basis of Maxwell's equations. In a solid, a similar coupling was first considered by Pierre Curie \cite{Curie1894} between the magnetization M and electric polarization P. This magneto-electric
 (ME) effect was recently understood to be potentially important for applications because in information technology, it would allow magnetic information to be written electrically (with low energy consumption) and to be read magnetically.
The ME effect was demonstrated and studied in the 1960s in Russia \cite{Smolenskii82Venevtsev94} and since then, many so called 'multiferroic' materials have been identified \cite{Fiebig05Eerenstein06}. However, so far the magnitude and operating temperatures of any observed ME coupling have been too small for applications. In fact, the only known multiferroic
 material of potential practical interest is bismuth ferrite, $BiFeO_{3}$ which is actually antiferromagnetic below $T_N \approx 370$ \textdegree C \cite{Smolenskii82Venevtsev94} and ferroelectric with a high Curie temperature: $Tc \approx 820$ \textdegree C \cite{Smith68}. As a result, in recent years, there has been a resurgence in the research conducted on this material. Moreover, epitaxial strain in $BiFeO_3$ thin film has been
  described as a unique means of enhancing magnetic and ferroelectric properties \cite{Wang03}. It is actually unclear whether this is indeed the case and in order to clarify this point, the intrinsic properties of the bulk material need to be better understood. It is stunning that although $BiFeO_3$ has been extensively studied over the past 50
   years, some of its most basic properties are still not fully known. For instance, it is only in 2007 that its spontaneous polarisation at room-temperature has been measured to be in excess of $100 \mu C/cm^2$ \cite{LebeugleAPL07}. Moreover, in the bulk, the coupling between magnetic and ferroelectric orders has never been fully clarified. This property has only been measured in
    thin films \cite{Zhao06} very recently. This lack of accurate data stems from the difficulty in making high quality single crystals. We have recently been able to grow such single crystals below their ferroelectric Curie temperature using the flux technique \cite{LebeuglePRB07,LebeugleAPL07}. They are usually produced in the form of platelets 40-50 microns thick and up to $3mm^2$ in area. Polarised light imaging and P(E) measurements \cite{LebeuglePRB07} indicate that the as-grown crystals are generally in a single ferroelectric/ferroelastic domain state. We report here on a neutron study of the coupling between magnetic and ferroelectric orders in two of these crystals.

\begin{figure}[ht]
\includegraphics[scale=0.6,draft=false,clip=true]{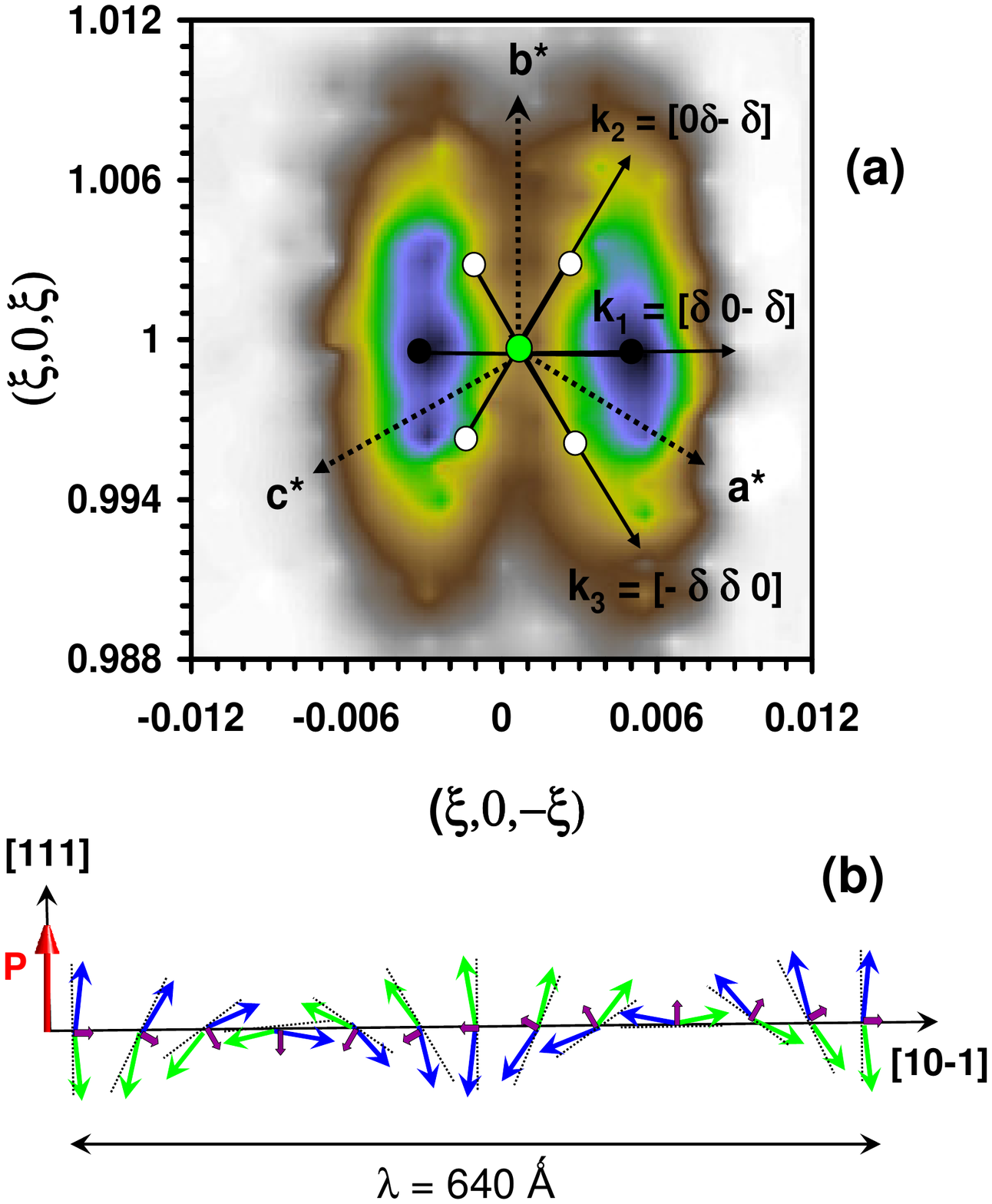}
\caption{(a) Neutron intensity around the $(\frac{1}{2},\frac{-1}{2},\frac{1}{2})$ Bragg reflection in the single domain state. The two diffraction satellites indicate that the cycloid is along the (10-1) direction. (b) Schematics of the magnetic configuration of antiferromagnetic vectors in the $64nm$ periodic circular cycloid.} \label{fig1}
\end{figure}

Our $BiFeO_3$ single crystals are rhombohedral at room temperature with the space group R3c and a pseudo-cubic cell with $a_{pc}= 3.9581$ \AA, $\alpha_{pc} = 89.375$\textdegree ($a_{hex}=5.567(8)$ \AA, $c_{hex} = 13.86(5)$ \AA  in the hexagonal
 setting), in perfect agreement with previous reported data \cite{Kubel90}. No ferroelastic twinning was observed and the elongated rhombohedral direction, which is parallel to the polarisation, is indexed as (111). $Fe^{3+}$ ions are ordered antiferromagnetically (G-type) and their moments describe a cycloid with a period of
$62 nm$, as has been established by early neutron diffraction data on sintered samples \cite{Sosnowska82,Przenioslo06}. Because of the rhombohedral symmetry, there are three equivalent propagation vectors for the cycloidal rotation: $\vec{k_1}=(\delta 0 -\delta)$,
$\vec{k_2}=(0 -\delta \delta)$ and $\vec{k_3}= (-\delta \delta 0)$ where $\delta = 0.0065$. In powder neutron diffraction, the different equally populated $\vec{k}$ domains lead to a splitting of magnetic peaks along three directions. Thus, as has been pointed out recently \cite{Przenioslo06}, the determination of modulated magnetic ordering is not unique because elliptical cycloids and Spin Density Waves (SDW), give the same diffraction pattern. The exact nature of the periodic structure is an important parameter for antiferromagnetic ferroelectrics since recent models of magnetoelectric coupling give a non-vanishing electric polarization for cycloids and elliptic ordering and zero
polarization for a SDW \cite{Przenioslo06,Mostovoy06}. It is possible to eliminate this ambiguity by measuring high-resolution scans around the strongest magnetic reflections of a single crystal. This is however a difficult experimental challenge because the long period imposes an
extremely high angular resolution. The diffractometre used in this work is 'Super 6T2' in the 'Laboratoire L{\'e}on Brilloin' in Saclay (France), where a resolution of $0.15$\textdegree vertically and $0.1$\textdegree horizontally can be achieved. We have measured the intensity distribution of the as-grown crystals around the four antiferromagnetic Bragg reflections
$(\frac{1}{2},\frac{-1}{2},\frac{1}{2})$, $(\frac{1}{2},\frac{1}{2},\frac{1}{2})$, $(\frac{1}{2},\frac{1}{2},\frac{-1}{2})$ and $(\frac{-1}{2},\frac{1}{2},\frac{1}{2})$. The peak splitting only occurs along one of the three symmetry allowed directions as shown for the $(\frac{1}{2},\frac{1}{2},\frac{1}{2})$ reflection in Fig. \ref{fig1}-a. Therefore, the modulated structure has a unique propagation vector $\vec{k_1}=(\delta 0 -\delta)$ with $\delta=0.0064(1)$ corresponding to a period of $64 nm$. The elongated shape of the measured sattelites is due to the better resolution in the horizontal direction (along (10-1)) but also possibly because of a slight warping of the sample induced by the silver epoxy electrodes apposed on both sides of the crystals (corresponding to the (010) plane) for electric poling.	

The spin rotation plane can also be determined because the magnetic scattering amplitude depends on the component of magnetic moments perpendicular to the scattering vector. A quantitative analysis of the integrated intensities of 10 theta/two-theta magnetic reflections (see table \ref{table1}) allows us to conclude unambiguously that the moments lie
in the plane defined by $\vec{k_1}=(\delta 0 -\delta)$ and the polarisation vector $\vec{P}//[111]$ (fig. \ref{fig1}-b). 

\begin{table}[ht]
	\centering
\begin{tabular}{|c|c|c|c|c|}
\hline
{\bf Bragg peak}  & {\bf $\vec{P}, \vec{k_1}$}  & {\bf $\vec{P}, \vec{k_2}$}  & {\bf $\vec{P}, \vec{k_3}$}  & {\bf $I_{obs}$} \\
\hline
{\bf $(\frac{1}{2}, \frac{-1}{2}, \frac{1}{2})$}  &  {\bf 189 }  &        122  &        122  & {\bf 198(8)} \\
\hline
{\bf $(\frac{1}{2},  \frac{1}{2}, \frac{1}{2}) $}  &  {\bf 100 }  &        100  &        100  & {\bf 99(6)} \\
\hline
{\bf $(\frac{-1}{2}, \frac{1}{2}, \frac{1}{2})$}  &  {\bf 122 }  &        122  &        189  & {\bf 116(6)} \\
\hline
{\bf $(\frac{1}{2}, \frac{1}{2}, \frac{-1}{2})$}  &  {\bf 122 }  &        189  &        122  & {\bf 114(6)} \\
\hline
\hline
{\bf $(\frac{1}{2}, \frac{-3}{2}, \frac{1}{2})$}  &  {\bf 113 }  &         71  &         71  & {\bf 111(11)} \\
\hline
{\bf $(\frac{3}{2}, \frac{-1}{2}, \frac{-1}{2})$}  &   {\bf 71  }  &         71  &        113  & {\bf 83(8)} \\
\hline
{\bf $(\frac{-1}{2}, \frac{-1}{2}, \frac{3}{2})$}  &   {\bf 71  }  &        113  &         71  & {\bf 83(8)} \\
\hline
{\bf $(\frac{-1}{2}, \frac{-3}{2}, \frac{-1}{2})$}  &   {\bf 71  }  &         61  &         61  & {\bf 78(9)} \\
\hline
{\bf $(\frac{3}{2}, \frac{1}{2}, \frac{1}{2})$}  &   {\bf 61  }  &         61  &         71  & {\bf 52(6)} \\
\hline
{\bf $(\frac{1}{2}, \frac{1}{2}, \frac{1}{2})$}  &   {\bf 61  }  &         71  &         61  & {\bf 56(7)} \\
\hline
\end{tabular}  
	\caption{Intensity measured around the magnetic Bragg positions compared to that expected for a cycloid with magnetic vectors in the different allowed ($\vec{P},\vec{k}$) planes.}
	\label{table1}
\end{table}
	
The structure refinement also confirms that the periodic structure is indeed a circular cycloid with antiferromagnetic moments  $\mu_{Fe} = 4.11(15) \mu_{B}$. Using a SDW model, or introducing a $20 \%$ ellipticity, deteriorates significantly the
 agreement factor of the fit. A consequence of the single $\vec{k}$ vector of the cycloid is that the crystal symmetry is lowered. Indeed, the ternary axis is lost and the average symmetry becomes monoclinic with the principal direction along $\vec{k}$=(110) \cite{SchmidPrivateComm}.

\begin{figure}[ht]
\includegraphics[scale=0.6,draft=false,clip=true]{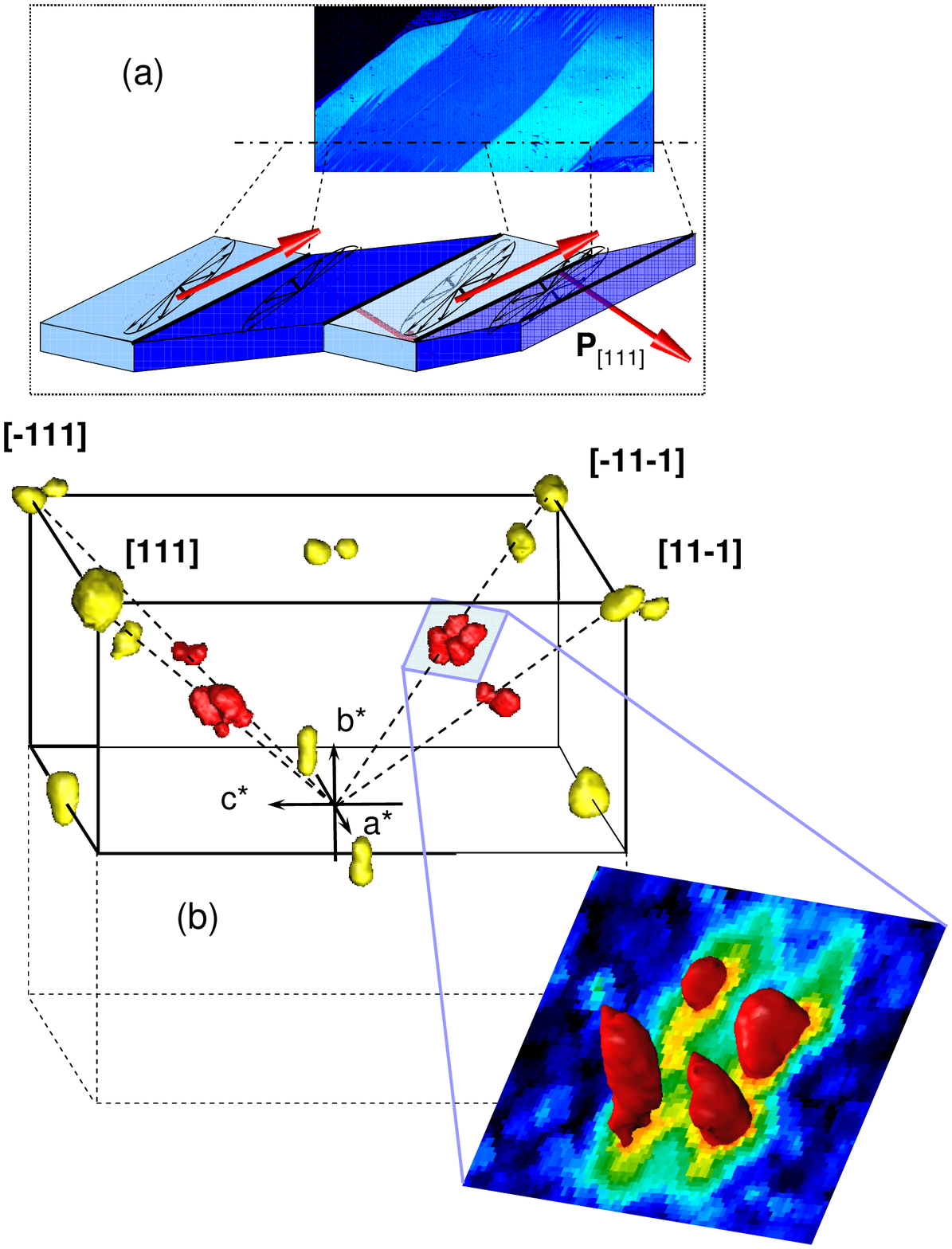}
\caption{Mapping of the neutron intensity in reciprocal space. Two sets of splitting appear for the nuclear intensity (yellow spots) consistent with the presence of two ferroelastic domains (see (a)): one because of the presence of two rhombohedral distortions along [111] and [1-11], and the second because of a physical buckling of the crystal induced by the twinning. Magnetic peaks are further split because of the cycloids. Note that because the splitting is small, the scale has been magnified by a factor of 10 on each peak position.} \label{fig2}
\end{figure}

No electric field effect on the magnetic order has ever been reported in bulk $BiFeO_3$. Here, we analyse the effect of poling in the (010) direction perpendicular to the platelet. Because polarisation changes force charges to re-organise, the polarisation state of the sample can be monitored by measuring the current in the circuit. As the first coercive field was reached, we measured a significant decrease in the
 $(\frac{1}{2},\frac{-1}{2},\frac{1}{2})$ neutron reflection intensity, indicating a redistribution of the average rhombohedral distortion. After reaching a multidomain state with $<P>\approx 0$, several neutron diffraction scans were performed. When trying to map precisely the intensity distribution around the antiferomagnetic Bragg
  positions, we found that the vertical resolution used for the crystal in its virgin state (in fig. \ref{fig1}) was not sufficient. Indeed, in the multi-domain state, ferroelastic distortions twin the crystal and complicate the diffraction patterns. In order to obtain a meaningful measurement, we had to reach 0.1\textdegree of resolution in both horizontal and vertical directions, which pushes the experimental conditions to the limit of what can be done with
   these instruments. Figure \ref{fig2}-b shows the (3D) reciprocal space mapping of the crystal. Yellow (111) type reflections are purely nuclear in origin while the red $(\frac{1}{2},\frac{1}{2},\frac{1}{2})$ are purely magnetic. (111) and (1-11) reflections are split along the long diagonals (dashed lines), which indicates the presence of two domains with different reticular distances. These are two rhombohedral twins with polarization axes along (111) and (1-11), roughly $50\%-50\%$ in volume. The
    other (-111) and (11-1) reflections are also split, but along the (101) direction. This is due to a buckling of the crystal schematically shown in fig. \ref{fig2}-a, which slightly changes the angles fulfilling the Bragg conditions. This is fully consistent with polarised optical microscope images taken on similar crystals (fig. \ref{fig2}-a) indicating that the multi-domain state consists of stripe regions with two different polarisation directions. 

\begin{figure}[ht]
\includegraphics*[scale=0.6,draft=false,clip=true]{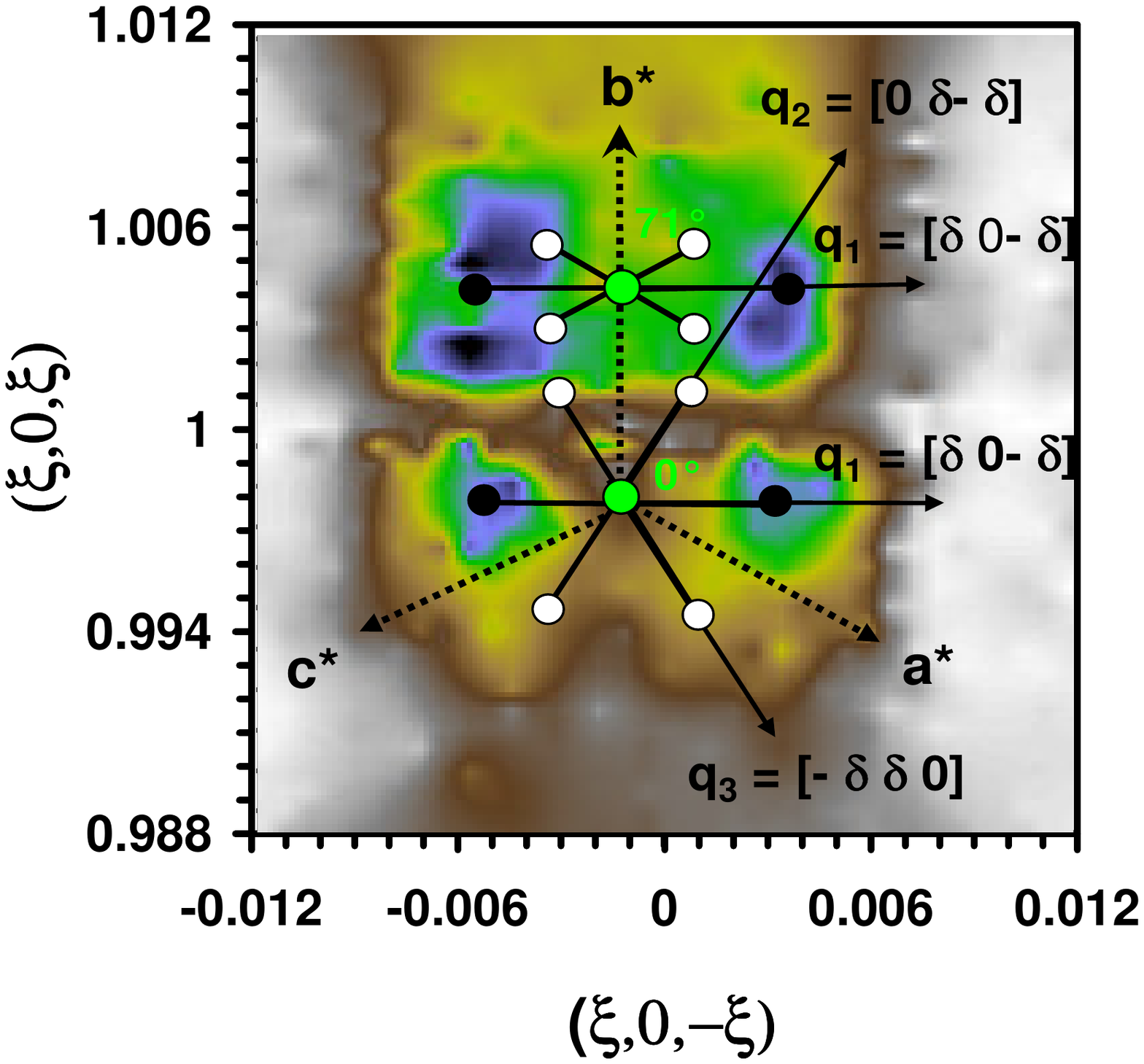}
\caption{Neutron intensity around the $(\frac{1}{2},\frac{-1}{2},\frac{1}{2})$ Bragg position in the multidomain state. Theoretical positions are indicated by the black and white spots. Diffraction satellites are visible in the 0\textdegree (bottom half) and 71\textdegree (top half) domains of polarisation. The difference in vertical spot shape likely originates from the position of reversed domains at opposite ends of the sample because of a prefered nucleation near the the edges. Any warping of the sample splits the new (1-11) peak while recovering an improved resolution for the original domain located near the center.} \label{fig3}
\end{figure}

The purely antiferromagnetic peaks have been analysed in more details. The strongest $(\frac{1}{2},\frac{-1}{2},\frac{1}{2})$ reflection is shown in the zoomed region of fig. \ref{fig2}-b to be composed of four spots. These result from two simultaneous splits, one due to the ferroelectric distortion (already evidenced in the nuclear peaks) and one of magnetic origin. A projection of the zoomed area is represented in figure \ref{fig3} on with the expected reflections from $P_{111}$ and $P_{1-11}$ domains are shown as green spots. The magnetic satellites are also indicated as black spots for the cycloid in the original (-101) direction and white spots for the other two symmetry allowed ones. The domains with the
  original $P_{111}$ direction of polarisation lead to the pattern in the lower half of the figure, while those where the polarisation rotated by 71\textdegree ($P_{1-11}$) are in the upper half. In the latter domain, the expected satellites are not in a regular rhombohedral symmetry because they do not belong to the (111) diffraction plane of the figure. However, were they present, these satellites would still appear because the figure is a projection. Figure.\ref{fig3} shows that in both domains, the splitting is only in the horizontal direction. Hence, the original $(\delta 0-\delta)$ propagation direction of the cycloid in the virgin state was retained everywhere.

\begin{figure}[ht]
\includegraphics*[scale=0.7,draft=false,clip=true]{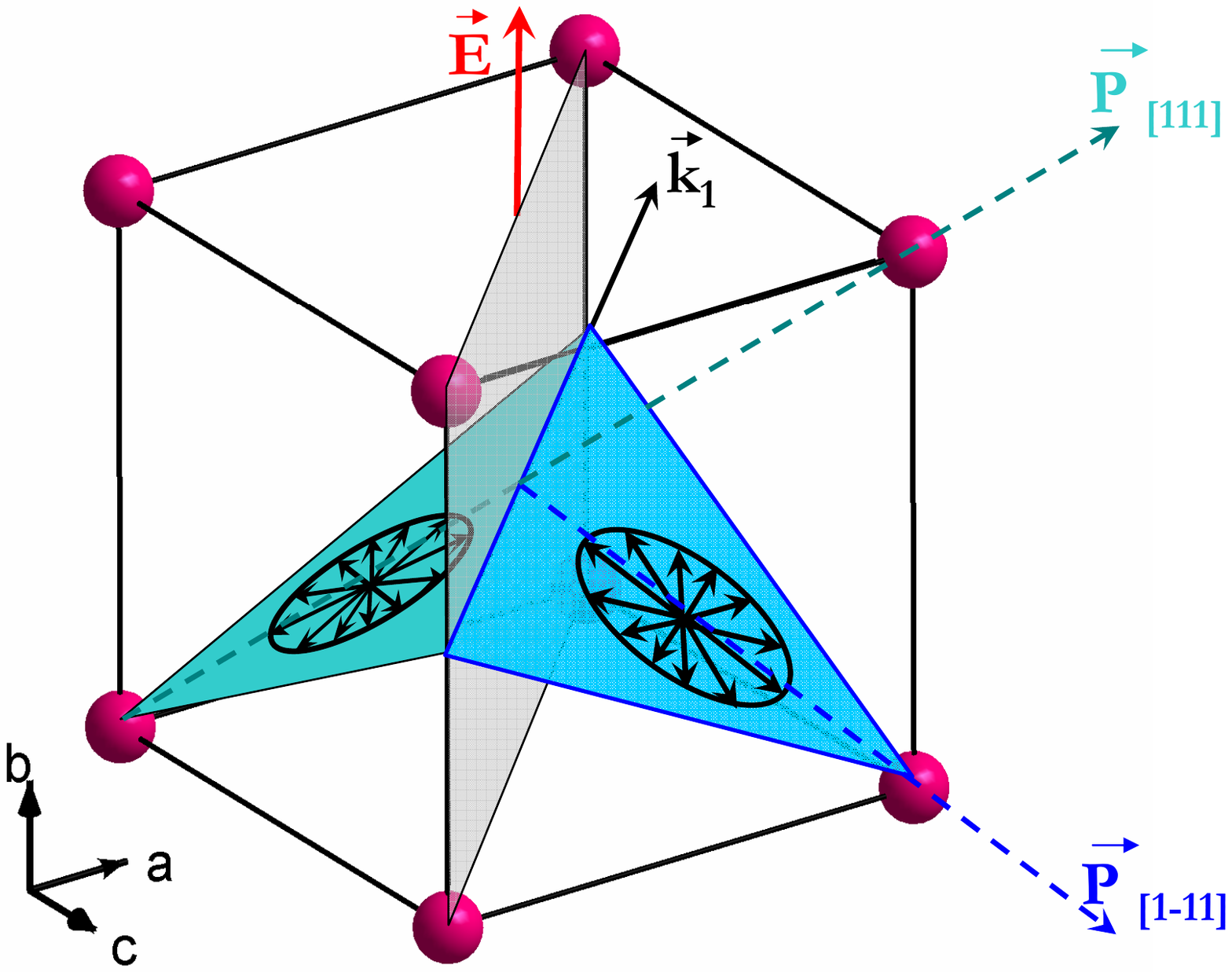}
\caption{Schematics of the planes of spin rotations and cycloids $\vec{k_{1}}$ vector for the two polarisation domains separated by a domain wall (in grey).} \label{fig4}
\end{figure}

The rotation planes of the AF vectors in the two domains can again be determined using the integrated intensities of the magnetic reflections (table \ref{table2}). These can be well accounted for by considering that $55\%$ of the crystal volume has switched its polarisation by 71\textdegree, and brought with it the rotation plane of the Fe moments. Thus, in each domain, AF moments are rotating in the plane defined by $\vec{k_1}$ and $\vec{P}$ as represented in fig.\ref{fig4}. Hence, the electric field induced change of polarisation direction produces a spin flop of the antiferromagnetic sublattice. 
 
\begin{table}[ht]
	\centering
\begin{tabular}{|c|c|c|c|c|c|c|}
\hline
{\bf Bragg peak}  & {\bf $\vec{P_{0}},\vec{k_1}$} & {\bf $\vec{P_{71}},\vec{k'_1}$} & {\bf $\vec{P_{71}},\vec{k'_2}$}  & {\bf $\vec{P_{71}},\vec{k'_3}$}  & {\bf $I_{obs}$} & {\bf $I_{calc}$} \\
\hline
{\bf $(\frac{1}{2}, \frac{-1}{2}, \frac{1}{2})$} &  {\bf 189} &  {\bf 100} &    100 &     100 & {\bf 158(7)} &  {\bf 150} \\
\hline
{\bf $(\frac{1}{2},  \frac{1}{2}, \frac{1}{2}) $}  &  {\bf 100} &  {\bf 189} &    122 &     122 & {\bf 145(6)} &  {\bf 139} \\
\hline
{\bf $(\frac{-1}{2}, \frac{1}{2}, \frac{1}{2})$}  &  {\bf 122} &  {\bf 122} &    122 &     189 & {\bf 120(8)} &  {\bf 122} \\
\hline
{\bf $(\frac{1}{2}, \frac{1}{2}, \frac{-1}{2})$} &  {\bf 122} &  {\bf 122} &    189 &     122 & {\bf 112(9)} &  {\bf 122} \\
\hline
\end{tabular}  
\caption{Intensity measured around the magnetic Bragg positions compared to that expected for a cycloid with magnetic vectors in the different allowed planes. Calculated values are obtained with $55\%$ of domains having switched their polarisation by 71\textdegree and kept the same propagation vector.}
	\label{table2}
\end{table}

These measurements unambiguously demonstrate that the magnetic $Fe^{3+}$ moments are intimately linked to the polarisation vector. This negates the common belief that in bulk $BiFeO_3$ magneto-electric coupling must be weak because the cycloid cancels linear ME effects \cite{Tabares85,Popov93,Scott94,Kadomtseva04}. Although $\left\langle M\right\rangle=0$ imposes a zero \textbf{global} linear ME effect in the bulk, the atomic coupling between $\vec{M}$ and $\vec{P}$ still exists. The underlying relevant microscopic mechanism is the (generalised) Dzyaloshinskii-Moriya (DM) \cite{Dzyaloshinskii59Moriya60} interaction which has recently been re-addressed starting from electronic Hamiltonians including spin-orbit coupling \cite{Katsura05,Sergienko06}. Katsura et al. \cite{Katsura05} describe in terms of spin currents the polarisation induced by a cycloidal spin arrangement, which can be written as $\vec{P} \propto \vec{e_{ij}} \times (\vec{S_i} \times \vec{S_j})$, with $\vec{S_{i,j}}$ the
   local spins and $\vec{e_{ij}}$ the unit vector connecting the two sites. The interaction of this polarisation with a coexisting internal polarisation produces a magneto-electric term in the total energy \cite{Katsura05}: $E_{DM} =(\vec{P} \times \vec{e_{ij}}).(\vec{S_i} \times \vec{S_j})$. This ME interaction, which can also be obtained from symmetry considerations \cite{Mostovoy06,Kadomtseva04}, was held responsible for the cycloidal spin arrangement in $BiFeO_3$ \cite{Kadomtseva04}. This coupling energy induces the canting of Fe moments which exactly compensates for the loss in exchange energy (neglecting the anisotropy energy): $E=-Ak^2$. A ME energy density of $-3.10^7 J/m^3$ can be inferred from the value of the period of the cycloid and the exchange constant ($A=3.10^{-6} J/m$). Importantly, the coupling energy is zero when $\vec{P}$ is perpendicular to the local moments and maximum when it lies in the cycloid rotation plane. This explains the
 antiferromagnetic flop we observe when $\vec{P}$ changes direction. This also explains why the two crystals we measured had their cycloids in the same direction $\vec{k_1}$. Indeed, this minimises the components of the magnetic spins parallel to the depolarisation field (normal to the platelets surface), which lowers the cost in DM energy.
 
When in thin film form, $BiFeO_3$ is a very different system because epitaxial strain suppresses the cycloid and induces a weak magnetic moment \cite{BeaPhilMag07Bea05}. Locally, the magnetic structure consists of canted spins with angles changing sign between neighbours, which makes the moments add. If this magnetic configuration were to generate a local polarisation, its direction would
  alternate from site to site. Therefore, in order for a global polarisation to coexist with weak ferromagnetism, it is better if the spins are in a plane perpendicular to the polarisation direction, a configuration for which the DM based interactions are zero. This is exactly what is observed in $BiFeO_3$ films \cite{Wang03}. The magnitude of this ME coupling is more difficult to estimate than that in the bulk, but because it
   originates from the frustration of the DM interactions, it is likely to be weaker. Interestingly, canting angles are only about 0.2\textdegree (the weak ferromagnetic moment being $0.02 \mu_{B}/atom$ \cite{BeaPhilMag07Bea05}), to be compared with 2.25\textdegree in the bulk. This underlines the ME origin of the cycloid and also hints at a stronger coupling in the bulk since the interaction between $\vec{P}$ and $\vec{M}$ is directly linked to canting.

In order to make a useful device with $BiFeO_3$, we suggest to use the exchange bias interaction between a thin ferromagnetic layer and a $BiFeO_3$ substrate, i.e. with its cycloid. It should be possible to vary electrically the exchange bias interaction using the antiferromagnetic flop observed here. Indeed, it is known that in conventional exchange bias systems, a non-compensated antiferromagnetic surface is not a prerequisite to obtain a large exchange field. Hence, it is likely that the cycloid may not significantly affect the bias, while optimising the coupling between the antiferromagnetic and ferroelectric orders.

We would like to aknowledge fruitfull discussions with Hans Schmid, Alexander Zhdanov and Daniil Khomskii as well as funding from the 'Agence Nationale de la Recherche' through the contract 'FEMMES'. We also thank X. Le Goff for X-ray measurements on single crystals as well as Gustau Catalan, Neil Mathur and Manuel Bib{\`e}s for their critical reading of the manuscript.



\end{document}